\documentclass[a4paper]{article}
\usepackage{graphicx}
\usepackage{subfigure}

\begin{document}

\title{\bf Memristive circuits and their Newtonian models $\phi ''=F(t,\phi,\phi')/m$ with memory}

\author{Wieslaw Marszalek\\
\small Rutgers University, 
 Department of Mathematics \\
\small 110 Frelinghuysen Rd., 
Piscataway, NJ 08854-8019, USA\\
 \small w.marszalek@rutgers.edu\\
       \vspace{.15cm}\\
        Zdzislaw Trzaska\\
\small Warsaw University of Ecology and Management\\
\small  12 Olszewska Str., 00-792 Warsaw, Poland} 
\maketitle

\begin{abstract}
The prediction made by L. O. Chua 45+ years ago (see: IEEE Trans. Circuit Theory (1971) 18:507-519  
and also: Proc. IEEE (2012) 100:1920-1927) about the existence of a passive
circuit element (called memristor) that links the charge and flux variables has been confirmed by the
HP lab group in its report (see: Nature (2008) 453:80-83) on a successful construction of such
an element. This sparked an enormous interest in mem-elements, analysis of their unusual
dynamical properties (i.e. pinched hysteresis loops, memory effects, etc.) and construction of
their emulators. Such topics are also of interest in mechanical engineering where memdampers (or memory dampers) play the role equivalent to memristors in electronic circuits. 
In this paper we discuss certain properties of the oscillatory memristive circuits, including those with mixed-mode oscillations.
Mathematical models of such circuits can be linked to the Newton's law $\phi''\!-\!F(t,\phi,\phi')/m=0$, 
with $\phi$ denoting the flux or charge variables, $m$ is a positive constant and
the nonlinear non-autonomous function $F(t,\phi,\phi')$ contains memory terms. This leads further to scalar fourth-order
ODEs called the jounce Newtonian equations. The jounce equations are used to construct the
$RC$+op-amp simulation circuits in SPICE. Also, the linear parallel $G$-$C$ and series $R$-$L$ circuits with sinusoidal
inputs are derived to match the \emph{rms} values of the memristive periodic circuits. 

\vspace{.15cm}

\vspace{.15cm}
\end{abstract}

\section{Mem-elements: action, coaction and one-period loops}
\label{intro}
\begin{figure}[h!]
\begin{center}
\includegraphics*[height=2.0in,width=2.6in]{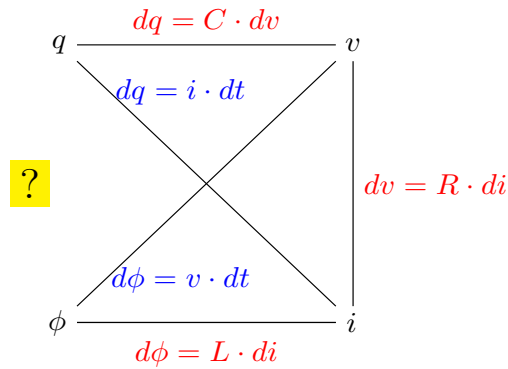}
\caption{The missing fourth element.}
\label{Fourth_Element} 
\end{center}
\end{figure}

\paragraph{Historical perspective: the fourth missing element} In 1971 L. O. Chua  predicted  existence of a passive circuit element that links the flux and charge variables \cite{paper1}. The missing element marked by the question mark in Fig.\ref{Fourth_Element} completes, together with the three well-known other passive elements (resistor, inductor and capacitor), the fourth side of the square diagram \cite{paper2}. In total, there are six relationships between the four variables of voltage $v$, current $i$, charge $q$ and flux $\phi$ as the two diagonal relationships are the well-known time-derivatives. It was not until 2008 when a group of researchers at the Hewllet-Packard lab has announced: \textit{'the missing element has been found'} \cite{paper3}. After the announcement a rather large number of results dealing with memristors have been reported in the literature (cf. [4-10] and references therein). Memristors, and also memcapacitors and meminductors (commonly referred to  as mem-elements) [6,11], are intriguing circuit elements not only from the point of view of circuit design, but also,  because of their unusual dynamical properties and characteristics. For that reason mem-elements attract interests of dynamical systems analysts, mathematicians (including numerical researchers) and computer engineers. Last, but not least, mechanical memristors are becoming important elements in mechanical and electro-mechanical devices [12,13] and memristive fingerprints occur in electro-mechanical Cassie-Mayr welding arcs models [14,15].

This paper further expands the notion of equivalence between electrical and mechanical elements and devices, as it provides explicit formulae for the \emph{force} quantity $F$ in $\phi ''(t)=F(t,\phi,\phi')/m$, the second Newton's law, for memristive Chua's circuits and other similar circuits with mixed-mode oscillations.

In general, there are six mem-elements, classified according to the input-output relationship and the nature of the \textit{internal} variable. All six mem-elements can be described as follows. 

Consider the $x$-controlled mem-element  described by the following equations
\begin{equation}\label{eq1}
y(t)=g(w(t))x(t),\hspace{.5cm}w'(t)=x(t) 
\end{equation}
where $y$ and $x$ are the  mem-element's output and input variables, respectively, $w$ is the internal state variable and the prime $'=d/dt$. For the six possible mem-elements of interest,
 the variables $y$, $x$ and $w$ have the meanings shown in table \ref{tab1}, where $v$, $i$, $q$, $\phi$, $TIQ$ and $TIF$ denote the voltage,
 current, charge, flux, time-integral of charge and time-integral of flux, respectively. The four letter abbreviations in the first column of table \ref{tab1} indicate the input variable and the type of mem-element. For example, the CCMR stands for a current controlled memristor. Since the input (controlling) variable $x$ is the current ($x=i$), therefore the internal variable $w$ is the charge, since $q'=i$ (see  (\ref{eq1})). The remaining output variable $y$ of CCMR  is the voltage $v$, thus $y=v$. The other five mem-elements are also described by (\ref{eq1}) with $MC$ and $ML$ in $XCMC$ and $XCML$ denoting memcapacitors and meminductors, respectively. The $X$ denotes the controlling (input) variable. Thus, depending on a mem-element, the $X$ is one of the variables from the set $\{Q,V,F,C\}$, either the charge, voltage, flux or current, respectively. Table \ref{tab1} includes the $g(w)$ function for each mem-element and the function $G(w)=\int_{\Gamma }g(w)dw=\int_{t_1}^{t_2}G(w)w'dt$, $G(0)=0$, for $t_1\le t\le t_2$ and $w(t_1)=w_1\in \Gamma$, $w(t_2)=w_2\in \Gamma$. The function $G(w)$ is used in the next section in the analysis of the \textit{action} parameter for mem-elements.  We assume that $g(w)$ are smooth functions of the internal variable $w$. Typical cases considered in the literature involve polynomials $g(w)=\sum_{i=0}^n a_iw^i$, $a_i\in R$ [2,3].

\begin{table}
\begin{center}
\caption{Six basic types of mem-elements}
\label{tab1}    
\begin{tabular}{llllll}
\hline \hline \noalign{\smallskip}
 & $ y $ & $ x $ & $ w$ & $ g(w)$ & $G(w)=\int_{\Gamma}g(w)dw$, [unit] \\
\noalign{\smallskip}\hline \hline \noalign{\smallskip}
\hspace{-0.15cm}VCMR & $i$ & $v$  & $\phi$ & $(MR(\phi))^{-1}$ & $\int_{\Gamma}(MR(\phi))^{-1}d\phi$, [A$\times$s]\\
\hspace{-0.15cm}CCMR & $v$ & $i$  & $q$ & $MR(q)$ & $\int_{\Gamma}MR(q)dq$,  [V$\times$s]\\
\noalign{\smallskip}\hline \noalign{\smallskip}
\hspace{-0.15cm}QCMC & $v$ & $q$ & $TIQ$ & $(MC(TIQ))^{-1}$& $\int_{\Gamma}(MC(TIQ))^{-1}d(TIQ)$,  [V$\times$s] \\
\hspace{-0.15cm}VCMC  & $q$ & $v$ & $\phi$  &$ MC(\phi)$   & $\int_{\Gamma}MC(\phi)d\phi$,  [A$\times s^2$] \\
\noalign{\smallskip}\hline \noalign{\smallskip}
\hspace{-0.15cm}FCML  &  $i$ & $\phi$ & $TIF$  & $(ML(TIF))^{-1}$ & $\int_{\Gamma}(ML(TIF))^{-1}d(TIF)$,  [A$\times$s] \\
\hspace{-0.15cm}CCML  &   $\phi$ & $i$ & $q$  & $ ML(q)$  & $ \int_{\Gamma}ML(q)dq$,  [V$\times s^2$] \\
\noalign{\smallskip}\hline \hline
\end{tabular}
\end{center}
\end{table} 

\paragraph{Action and coaction parameters} The action and coaction parameters [16-20] for memristors are discussed in details in [21], where both parameters are defined for all six mem-elements in the context of Euler-Lagrangian. For VCMR and CCMR the action $\mathcal{A}(t)$ and coaction $\hat{\mathcal{A}}(t)$ are defined as  $\mathcal{A}(t)\equiv \int _{w(0)}^{w(t)}G(w)dw$ and $\hat{\mathcal{A}}(t)\equiv \int _{G(w(0))}^{G(w(t))}wd(G(w))$, respectively. It follows from the definition of $\mathcal{A}(t)$ that $\mathcal{A}(t)=G(w)w|_0^t-\int_0^tg(w)ww'dt$. Moreover, it is also true that 
\begin{equation}\label{eq88}
\begin{array}{rl}\mathcal{A}(t)+\hat{\mathcal{A}}(t)=&\int_0^t\left (G(w)w'+w\frac{dG(w)}{dt}\right )dt\\
  =&\int_0^t\frac{d}{dt}\left (G(w)w\right )dt=G(w(t))w(t)-G(w(0))w(0).\end{array}
\end{equation}
which yields $\mathcal{A}(0)+\hat{\mathcal{A}}(0)=\mathcal{A}(T)+\hat{\mathcal{A}}(T)=0$ where $T$ denotes the period.

Analogous definition and properties hold true for MC and ML elements. In \cite{paper16} a detailed example with a derivation of the expression   $\mathcal{A}(0)+\hat{\mathcal{A}}(0)$ in terms of the polynomial coefficients of $g(w)$ and the internal variable $w$ has been reported. Also, \cite{paper16} includes a proposition to call the unit of action \emph{Chua} to honor L. O. Chua for his contribution in the area of memristors and memristive devices.

\paragraph{One-period loops} Suppose that we consider two periodic functions, $f(t)$ and $h(t)$ for $0\le t\le T$. Let $\Gamma$ denotes a loop in the $(f,h)$ plane and consider the quantities of the form 
  $\int_{\Gamma}\!fdh\!=\!\int_0^Tf(t)h'(t)dt$ (or $\int_{\Gamma}hdf\!=\!\int_0^Th(t)f'(t)dt$). In the context  of the six mem-elements  from Table \ref{tab1} and periodic functions $x$, $y=g(w)x$, $w$ and $G(w)$ one can define the six quantities based on six different pairs of periodic functions, as shown in Table \ref{tab2}.

\begin{table}
\begin{center}
\caption{Six pairs of quantities for each  mem-element in Table \ref{tab1}}
\label{tab2}     
\begin{tabular}{llll}
\hline\noalign{\smallskip}
$(f,h)$ & $ \int_{\Gamma}fdh$ & $\int_{\Gamma}hdf$ \\
\noalign{\smallskip}\hline\noalign{\smallskip}
$(g(w)x,x)$ & $\int_{\Gamma_1}g(w)xdx$ & $\int_{\Gamma_1}xd(g(w)x)$\\
$(x,G(w))$  & $\int_{\Gamma_2}xd(G(w))$ & $\int_{\Gamma_2}g(w)dx$\\
$(g(w)x,w)$ & $\int_{\Gamma_3}g(w)xdw$ & $\int_{\Gamma_3}wd(g(w)x)$ \\
$(G(w),w)$ & $\int_{\Gamma_4}G(w)dw$  & $\int_{\Gamma_4}wd(G(w))$\\
$(g(w)x,G(w))$ &  $\int_{\Gamma_5}g(w)xd(G(w))$ & $\int_{\Gamma_5} G(w)d(g(w)x)$\\
$(x,w)$ &   $\int_{\Gamma_6}xdw$ & $\int_{\Gamma_6}wdx$\\
\noalign{\smallskip}\hline
\end{tabular}
\end{center}
\end{table}

\begin{figure}[b!]
\begin{center}
\includegraphics*[height=2.0in,width=3in]{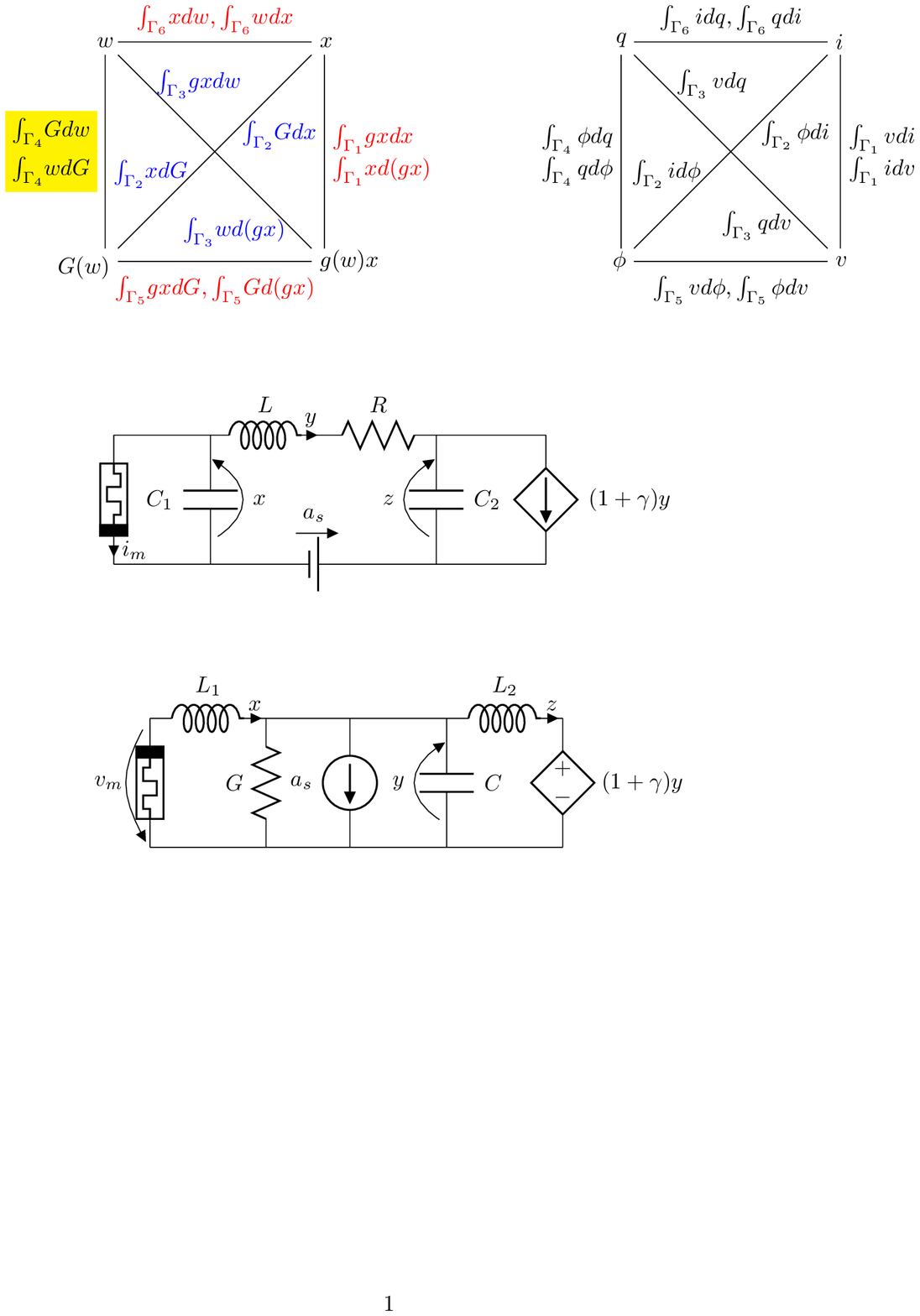}
\caption{Six pairs of quantities defined in Table \ref{tab2}.}
\label{Fig2} 
\end{center}
\end{figure}

Graphical representation of the integrals in Table \ref{tab2} is shown in Fig.\ref{Fig2}. Interpretation of the five quantities (integrals) along the top, bottom, right sides and two diagonals in Fig.\ref{Fig2} is given in \cite{paper16}. Notice that the two integrals on the left side of the square are the action and coaction parameters of the VCMR and CCMR. This shows a one-to-one correspondence of the diagram in Fig.\ref{Fig2} with that in Fig.\ref{Fourth_Element}. The quantities on the left side of the square in Fig.\ref{Fig2} for MC and ML elements are different than for the MR elements (see \cite{paper16}). 

\section{Oscillatory memristive circuits}

\begin{figure}[b!]
\begin{center}
\subfigure[Regular Chua's circuit]
{\label{Rys2a}\includegraphics*[height=0.9in,width=2.2in]{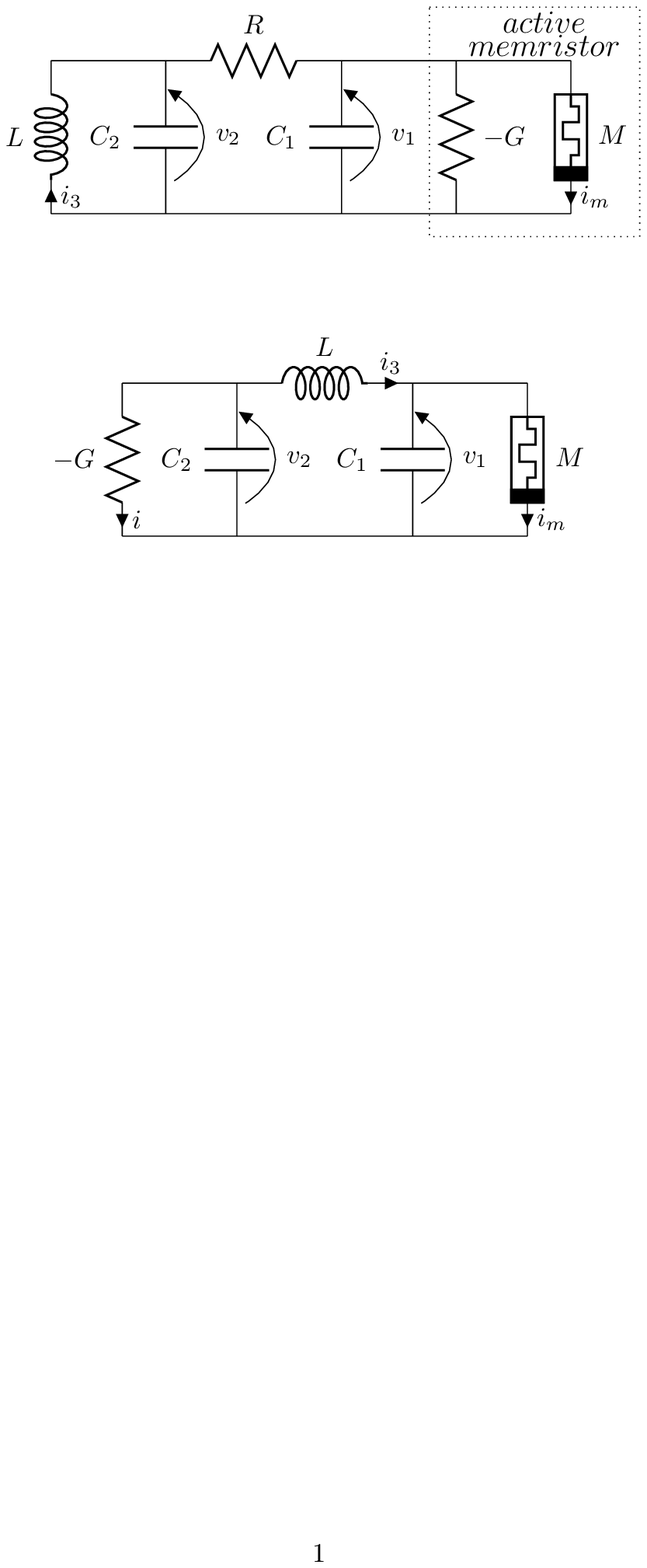}}
\subfigure[Canonical Chua's circuit]
{\label{Rys2b}\includegraphics*[height=0.9in,width=2.2in]{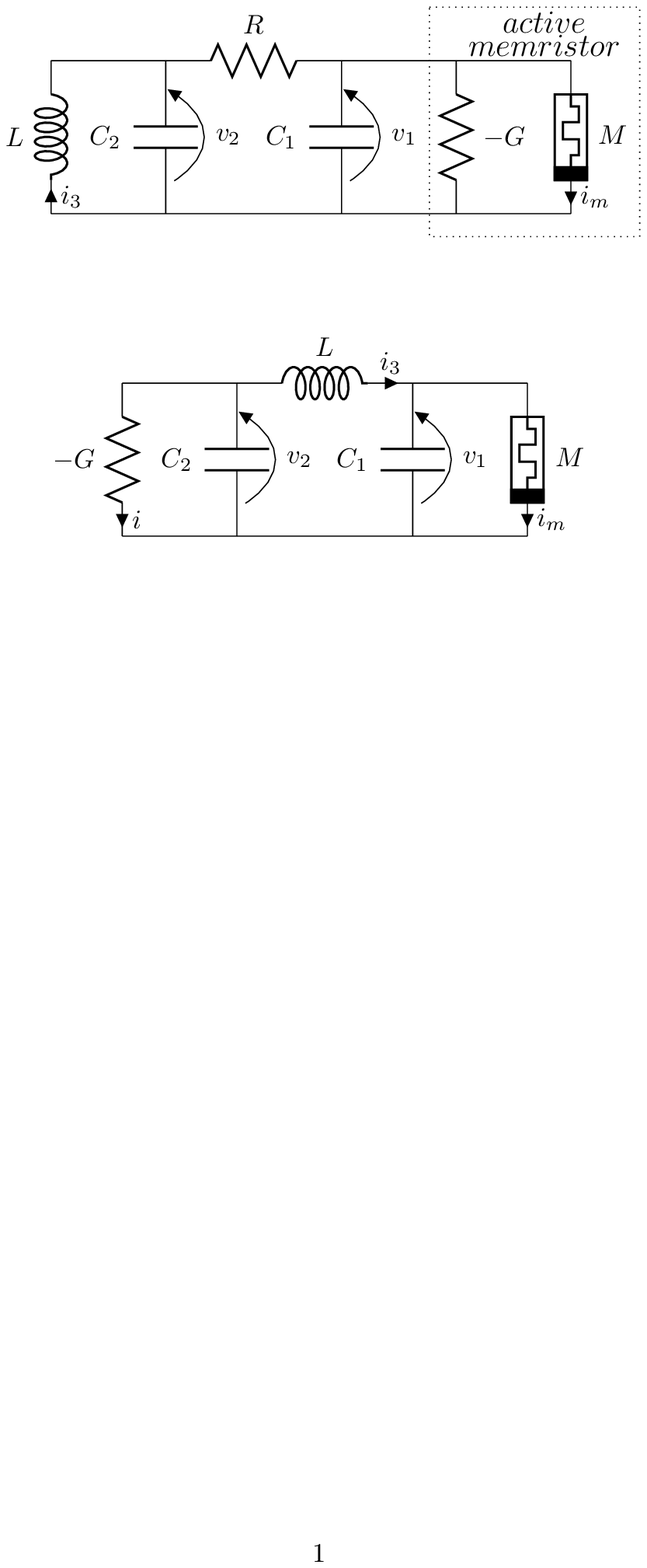}} 
\subfigure[Circuit with MMOs (ver. 1)]
{\label{Rys2c}\includegraphics*[height=0.9in,width=2.3in]{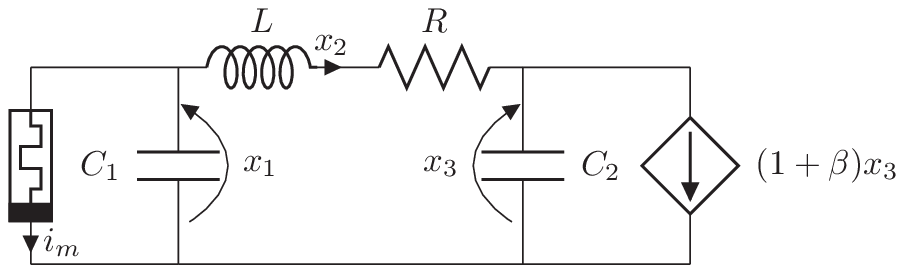}}
\subfigure[Circuit with MMOs (ver. 2)]
{\label{Rys2d}\includegraphics*[height=0.95in,width=2.3in]{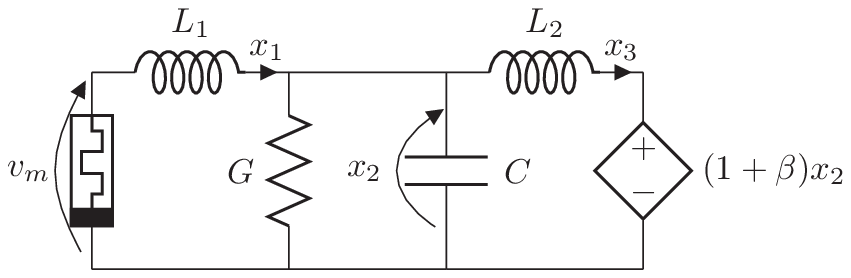}} 
\caption{Various oscillatory memristive circuits.}
\end{center}
\end{figure}

Figs.\ref{Rys2a}-\ref{Rys2d} show four typical oscillatory circuits with memristors. The first two circuits are the well-known Chua's regular and canonical circuits in which the piecewise-linear Chua's diodes have been replaced with memristors. Such circuits show various types of period-$n$ oscillations, where $n\in \{1,2,\dots \}$ depends on the circuits' parameters. The other two circuits (Figs.\ref{Rys2c} and \ref{Rys2d}) are oscillatory circuits with mixed-mode oscillations (or MMOs) of type $L^s$, where $L$ and $s$ denote the numbers of large and small amplitude oscillations in one period [20-27]. The circuits in Fig.\ref{Rys2c} and \ref{Rys2d} are dual in the sense that they are described by the same set of four first-order ODEs, namely
\begin{equation}\label{eq19}
\begin{array}{rcl} \epsilon \overline{x}' & = & s_c[-y/\eta -g(w)\overline{x}] \\
y' & = &s_c\alpha (\eta \overline{x}-Ky-z)\\
z' & = &-s_c\beta y\\
w' & = & s_c\eta \overline{x}\end{array}
\end{equation}

\begin{figure}[h!]
\begin{center}
\includegraphics*[height=3.3in,width=4.9in]{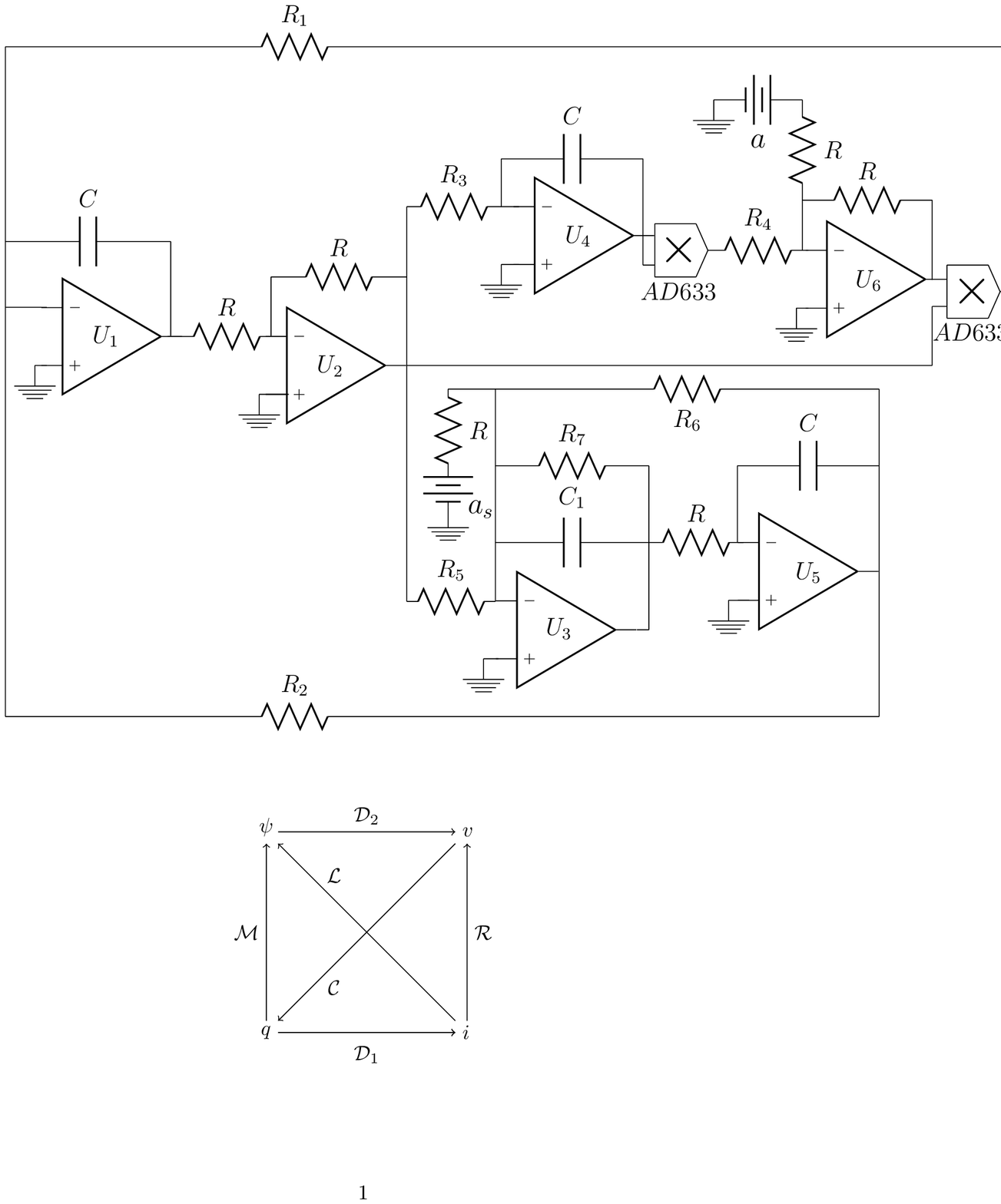}
\caption{Realization of dual memristive circuits with MMOs and $g(w)=a+3bw^2$.}
\label{Fig5} 
\end{center}
\end{figure}

\noindent where the prime $'$ denotes the time derivative, $0<C_1\equiv \epsilon \ll 1$, $\alpha=1/L$, $K=R$, $\beta =\gamma/C_2$ for the circuit in Fig.\ref{Rys2c} and $0<L_1\equiv \epsilon \ll 1$, $\alpha =1/C$, $K=G$, $\beta =\gamma/L_2$ for the circuit in Fig.\ref{Rys2d}. The current-controlled current source and voltage-controlled voltage source in the circuits are described through the expression $(1+\gamma)y$ with $\gamma>0$.  The scaling factor $\eta >1$ was chosen to reduce the variable $x$ and its derivative (important in circuit simulation in SPICE \cite{paper20}), since $\overline{x}=x/\eta$, with $x$ being the memductor's voltage in the circuit in Fig.\ref{Rys2c} and memristor's current in the circuit in Fig.\ref{Rys2d}. The $s_c>0$ is  a time scaling coefficient. In order to be able to excite the circuits from zero initial conditions, one can, for example, consider (\ref{eq19}) with the second equation replaced by $y'=s_c\alpha(\eta \overline{x}-Ky-z\pm a_s)$, that is, a small biasing constant source $a_s$ of order $\epsilon$ is used. The $a_s$ is a voltage source added to the $R$-$L$ branch in Fig.\ref{Rys2c} or a parallel current source added to the branches of $G$-$C$ in Fig.\ref{Rys2d}. When $a_s\ne 0$ one can use zero-initial conditions to obtain MMOs. Otherwise, with $a_s=0$, non-zero initial conditions should be used. Due to the \textit{small} values of capacitance $C_1$ and inductance $L_1$ in the circuits shown in Figs.\ref{Rys2c} and \ref{Rys2d}, respectively, the two circuits and their model (\ref{eq19}) are singularly perturbed ones.

A SPICE version of the circuits in Figs.\ref{Rys2c} and \ref{Rys2d} is shown in Fig.\ref{Fig5} with the assumption that  $g(w)=a+3bw^2$ in (\ref{eq1}). The output of the first multiplier AD633 gives $w^2(t)$, while the output of the second multiplier AD633 yields
 $(a+3bw^2)\overline{x}$. Also, $C_1=C\alpha/s_c$, $R_1=0.1\epsilon R/s_c$, $R_2=\eta \epsilon R/s_c$, $R_3=R/s_c$,
 $R_4=0.1R/(3b\eta ^2)$, $R_5=R/\eta$, $R_6=R/(\beta s_c)$, $R_7=R/K$, $V=a_s$ and $RC=1$ second.

\section{Newtonian properties and jounce equations for Chua's circuits}

We shall now examine the \emph{regular} and \emph{canonical} Chua's circuits with memristors shown in Figs.\ref{Rys2a} and \ref{Rys2b} and demonstrate that both circuits can be described by the second Newton's law from which jounce equations can be easily derived. We assume that inductor $L$ in Fig.\ref{Rys2a} has an internal resistance $r$. This circuit is described by the following equations 
\begin{equation}\label{eq33}
\begin{array}{rcl} x' & = & k\alpha[y+(\xi-1)x-g(w)x]\\
y' & = &k(x-y+z)\\
z' & = &-k(\beta y+\gamma z)\\
w' & = & kx.\end{array}
\end{equation}
where $[x,y,z,w]=[v_1,v_2,i_3,\phi ]$, $k>0$, $\alpha =1/C_1$, $\beta =1/L$, $\gamma=r/L$, $\xi=G$, $R=1$ and $C_2=1$. The circuit comprises the negative conductance $(-G<0)$, capacitors $C_1$ and $C_2$, resistor $R$, inductor $L$ (with an internal resistance $r$), a flux-controlled mem-element with memductance $g(w)$, where $w$ denotes the flux variable. The quantities $v_1$, $v_2$ and $i_3$ are the two voltages on $C_1$, $C_2$ and current through $L$, respectively.

\vspace{.2cm}
\noindent{\bf Theorem 1:}  The memristor's internal variable $w$ in the regular Chua's circuit satisfies $w''=F(t,w,w')/{m}$ with
\begin{equation}\label{tw1}\begin{array}{rl}
F/m=& \,\,k(\alpha h(w)-1-\gamma)w'+k^2(\alpha+\beta+\gamma)w-\alpha k^3\gamma\!\int\!wdt \\
&+\alpha k^2(1+\gamma)\!\int\!h(w)dw-\alpha k^3(\beta+\gamma)\!\int\!\int\!\!h(w)dwdt
\end{array}\end{equation}
and $h(w)=\xi-1-g(w)$. \hfill $\diamond$

\noindent {\emph{Proof.} The second and fourth equations in (\ref{eq33}) yield 
\begin{equation}\label{eq44} 
\begin{array}{rl}w'' &=\alpha k^2y+k\alpha h(w)w' \\
&=k\alpha h(w)w'+\alpha k^2 w-\alpha k^3\int ydt+\alpha k^3\int zdt. 
\end{array} \end{equation}
In order to express $\int ydt$ and $\int zdt$ in terms of $w$ and $w'$ we ntegrate the third equation in (\ref{eq33}) and substitute the integral of the second equation to get
$$z=-\beta\!\int\!kydt-\gamma\!\int\!kzdt=-(\beta+\gamma)\!\int\!kydt+\gamma w-\gamma y.$$
The first equation in (\ref{eq33}) gives $\int kydt=\frac1{\alpha k}w'-\int h(w)dw$. Therefore,
$\int zdt=-\frac{\beta+\gamma}{\alpha k}w-(\beta+\gamma)\int\int h(w)dwdt-\gamma\int wdt-\gamma\int ydt$, and  (\ref{eq44}) becomes
\begin{equation}\label{tw1_last}\begin{array}{rl}
w''&=k\alpha h(w)w'+\alpha k^2 w-\alpha k^3\int ydt-k^2(\beta+\gamma)w\\
&\hspace{.5cm}-\alpha k^3(\beta+\gamma)\int\int h(w)dwdt-\alpha k^3\gamma\int wdt-\alpha k^3\gamma\int ydt \\
&=k\alpha h(w)w'+k^2(\alpha+\beta+\gamma)w-\alpha k^2(1+\gamma)\int kydt \\
&\hspace{.5cm}-\alpha k^3\gamma\int wdt-\alpha k^3(\beta+\gamma)\int\int h(w)dwdt \\
&=k\alpha h(w)w'+k^2(\alpha+\beta+\gamma)w-k(1+\gamma)w'+\alpha k^2(1+\gamma)\int h(w)dw \\
&\hspace{.5cm}-\alpha k^3\gamma\int wdt-\alpha k^3(\beta+\gamma)\int\int h(w)dwdt.
\end{array}\end{equation}
The right side of (\ref{tw1_last}) is equal to the right side of (\ref{tw1}). \hfill $\diamond$

\vspace{.15cm}
The canonical Chua's circuit in Fig.\ref{Rys2b}  is described by the following equations \
\begin{equation}\label{eq68}
\begin{array}{rcl} x' & = & k\alpha[y-g(w)x] \\
y' & = &k(z-x)\\
z' & = &-k(\beta y+\gamma z)\\
w' & = & kx\end{array}
\end{equation}
where $[x,y,z,w]=[v_1,v_2,i_3,\phi ]$, $k>0$, $\alpha =1/C_1$, $\beta =1/C_2$, $\gamma=G/C_2$ and $L=1$. As before, the quantities $v_1$, $v_2$ and $i_3$ are the two voltages on $C_1$, $C_2$ and current through $L$, respectively.

\vspace{.2cm}
\noindent{\bf Theorem 2:} Variable $w$ in (\ref{eq68}) satisfies $w''=F(t,w,w')/m$ with $F/m$ given below.
 \hfill $\diamond$

\noindent \emph{Proof.} The first and fourth equations in (\ref{eq68}) yield
\begin{equation}\label{pr11}\begin{array}{rl}
w''=kx'=&k\alpha \left[ky-g(w)w'\right]
= k\alpha\left[-g(w)w'+k^2\int\left(z-x\right)dt\right] \\
 =& k\alpha\left[-g(w)w'-kw+k^2\int zdt\right].
\end{array}
\end{equation}
Variable $z$ satisfies the following equation
\begin{equation}\label{pr12}\begin{array}{rl}
z= k\int\left (-\beta y+\gamma z\right )dt=&-k\beta\int\left[\frac{1}{k\alpha}x'+g(w)x\right]dt+k\gamma\int\left(\frac{1}{k}y'+x\right)dt\\
 =&-\frac{\beta}{\alpha k}w'+\beta\int g(w)w'dt+\gamma\left[\frac{1}{k\alpha}x'+g(w)x\right]+\gamma w\\
 =& -\frac{\beta}{\alpha k}w'+\beta\int g(w)dw+\frac{\gamma}{k^2\alpha}w''\!+\!\frac{\gamma}{k}g(w)w'\!+\!\gamma w.
\end{array}
\end{equation}
Next, (\ref{pr12}) gives
\begin{equation}\label{pr13}
\int\!zdt=-\frac{\beta}{\alpha k}w+\!\beta\!\int\!\!\int \!g(w)dwdt+\frac{\gamma}{k^2\alpha}w'+\frac{\gamma}{k}\!\int\!g(w)dw+\gamma\!\int \!wdt.
\end{equation}
By using (\ref{pr13}) we obtain from (\ref{pr11})
\begin{equation}\label{pr14}\begin{array}{rl}
w''\!=\!&\!\!\!-k\alpha g(w)w'\!-\!k^2\alpha w\!-\!k^2\beta w\!+\!k^3\alpha \!\int \!\! \int \!\!g(w)dwdt\!+\!k\gamma w'\\
 &\hspace{1cm}+\,k^2\gamma \alpha \!\int \!\!g(w)dw\!+\!k^3\gamma \alpha \!\int \!\!wdt\\
 =&\!\!\!-k^2(\alpha\!+\!\beta)w+\!k\left[\gamma \!-\alpha g(w)\right]\!w'\!+k^3\gamma \alpha \!\int \!wdt+\!k^2\gamma \alpha \!\int \!g(w)dw\\
 &\hspace{1cm}+\,k^3\alpha \!\int \!\!\!\int g(w)dwdt\\
:=&\!\!\!F(t,w,w')/m.
\end{array}
\end{equation}
Thus (\ref{pr14}) yields $F/m$ which ends the proof. \hfill $\diamond$

\vspace{.2cm}
\noindent {\bf Corollary 1:} The $w$ in (\ref{eq68}) satisfies a jounce equation
\begin{equation}\label{pr15}\begin{array}{rl}
w''''+k\{\alpha g(w)-\gamma \}w'''+&\!\! k\{k\alpha-k\gamma \alpha g(w)+k\beta+3\alpha g'(w)w'\}w''\\
&\hspace{-1.5cm}+\,k^3\alpha\{\beta g(w)-\gamma \}w'\!-\!k^2\alpha \gamma g'(w)(w')^2\!+\!k\alpha g''(w)(w')^3=0.\end{array}
\end{equation}
 \hfill $\diamond$

\noindent \emph{Proof.}  Applying $\frac{d}{dt}$ twice to (\ref{pr14}) gives (\ref{pr15}). The corollary can also be proved in an alternative way. One differentiation of the last equation in (\ref{eq68}) together with the first equation yield 
\begin{equation}\label{eq3}
w''= k^2\alpha y - k^2\alpha g(w)w'.
\end{equation}
Next, from (\ref{eq3}) we obtain $y$, $y'$ and $y''$ as
\begin{equation}\label{eq4}\begin{array}{rcl}
y &=&w''/(k^2\alpha) +g(w)w'/k\\
y'&=&w'''/(k^2\alpha)+g'(w)(w')^2/k+g(w)w''/k\\
y''&=&w''''/(k^2\alpha)+g''(w)(w')^3/k+3g'(w)w'w''/k+g(w)w'''/k.\end{array}
\end{equation}
The second equation in (\ref{eq68})  can be written as $z=\frac{1}{k}y'+x$. Substituting it into the third equation in (\ref{eq68}) and using $x=w'/k$  we arrive at 
\begin{equation}\label{eq5}
y''+w''=-k^2\beta y+k\gamma y' + k\gamma w'.
\end{equation}
Now, inserting $y$, $y'$ and $y''$ from (\ref{eq5}) into (\ref{eq4}) yields (\ref{pr15}).  \hfill $\diamond$

\vspace{0.1cm}
It can also be shown that the variable $w$ in the regular Chua circuit satisfies a similiar jounce equation. The proof is omitted here.

\section{Oscillatory memristive circuits with MMOs and their jounce Newtonian properties}

Transformation $x=\eta \overline{x}$ and the second equation in (\ref{eq1}) yield the following (we use a nonzero $a_s$ as described in section 2).

\vspace{.2cm}
\noindent{\bf Theorem 3:} The $w$ in (\ref{eq1}) satisfies a Newtonian law $w''=F(t,w,w')/m$.
 \hfill $\diamond$

\noindent \emph{Proof.}  One differentiation of the last equation in (\ref{eq1}) and using the other equations in (\ref{eq1}) yield 
\begin{equation}\label{pr1}\begin{array}{rl}
w''=s_c x'=&\!\!\!\!-\frac{s_c}{\epsilon}\left[s_cy+g(w)w'\right]
=-\frac{s_c}{\epsilon}\left[g(w)w'+s_c^2\alpha\int\!\left(x-Ky-z\pm a_s\right)dt\right] \\
=&\!\!\!\!-\frac{s_c}{\epsilon}\left[g(w)w'+s\alpha w+\frac{s_c\alpha K}{\beta}z-s_c^2\alpha\int\!\! z(t)dt\pm s_c^2\alpha a_s t\right]. \end{array}
\end{equation}
In order to find $\int z(t)dt$ let's  determine $z$ first with 
\begin{equation}\label{pr2}\begin{array}{rl}
 z=&\!-s_c\beta\int\!ydt=_1\beta\int(\epsilon x'+g(w)w')dt \\
=&\!\frac{\beta\epsilon}{s_c}\int\!w''dt+\beta\int\!g(w)dw=\frac{\beta\epsilon}{s_c}w'+\beta\int\!g(w)dw.\end{array}
\end{equation}
This yields $\int\!zdt$ 
\begin{equation}\label{pr3}
\int\!zdt=\frac{\beta\epsilon}{s_c}w+\beta\!\int\!\!\int\! g(w)\,dwdt.
\end{equation}
Finally, we compute 
\begin{equation}\label{pr4}\begin{array}{rl}
 w''=\!\!\!\!& -\frac{s_c}{\epsilon}\{ s_c\alpha\!(1\!-\!\beta\epsilon)w\!+\!s_c\alpha K\!\!\int\!\!g(w)dw
\!-\!s_c^2\alpha\beta\!\!\int\!\!\int\!\!g(w)\!dwdt\!+\!(g(w)\!\\ &\hspace{1cm}+\!\alpha K\epsilon)w'\!\pm\! s_c^2\alpha a_s t\} \\
=\!\!\!\!& -\frac{s_c}{\epsilon}\{ \left(g(w)\!+\!\alpha K\epsilon\right)w'\!+\!s_c\alpha(1\!-\!\beta\epsilon)w\pm s_c^2\alpha a_st
\!+\!s_c\alpha K\!\!\int\!\!g(w)dw\\ &\hspace{1cm}-\!s_c^2\alpha\beta\!\!\int\!\!\int\!\! g(w)\!dwdt\} \\
:=\!\!\!\!& F(t,w,w')/m\end{array}
\end{equation}
which yields $F(t,w,w')/m$. \hfill $\diamond$

\vspace{.15cm}
\noindent \emph{Comment 3.1. } Due to the terms $\int\!\!\int\! g(w)dwdt$ (depending on $t$) and $s_c^2\alpha a t$,  the $F$ is non-autonomous.

\vspace{.2cm}
\noindent{\bf Corollary 2:} The system (\ref{eq1}) yields a jounce equation in $w$ in the form 
\begin{equation}\label{pr5}\begin{array}{rl}
\epsilon w''''+s_c\{\alpha K\epsilon+g(w)\}w'''+&\!\!\{s_c^2\alpha+s^2Kg(w)-s_c^2\alpha\beta\epsilon+3s_cg'(w)w'\}w''\\
&\hspace{-2cm}-s_c^3\alpha\beta g(w)w'+s_c^2\alpha Kg'(w)(w')^2+s_cg''(w)(w')^3=0.\end{array}
\end{equation}
 \hfill $\diamond$

\noindent \emph{Proof.} Applying $\frac{d}{dt}$ twice to (\ref{pr4}) yields (\ref{pr5}). \hfill $\diamond$

\vspace{.15cm}
Analogous results hold true for $x$, $y$ and $z$. We state the results below, but, to avoid unduly replications, the proofs will be omitted.

\vspace{.2cm}
\noindent{\bf Theorem 4:} $\hfill$

\vspace{.1cm}
\noindent {\bf (1)}  For a Newtonian formulation $x''=F(t,x,x')/m$ of (\ref{eq1}) we have
\begin{equation}\label{pr6}\begin{array}{rl}
\frac{F}m= 
  & -\frac{s_c}{\epsilon}\{ \pm s_c\alpha a_s\!-\!s_c^2\alpha\beta\!\int\!g(w)dw\!+\!(s_c\alpha\!-\!s_c\alpha\beta\epsilon\!+\!s_c\alpha Kg(w)\!+\!g'(w)x)x\\
  &+(\alpha K\epsilon\!+\!g(w)) x'\} \end{array}
\end{equation}
and $w(t)=s_c\int x(t)dt$.

\vspace{.1cm}
\noindent {\bf(2)} For a Newtonian formulation $y''=F(t,y,y')/m$ of (\ref{eq1}) we have 
\begin{equation}\label{pr7}\begin{array}{rl}
\frac{F}m=& \\
 &\hspace{-0.8cm}s_c+c\alpha\!\left(\!-\!\left[K\!+\frac{g(w)}{\epsilon\alpha}\right]\!y'+s_c\!\left[\beta-\frac1{\epsilon}-\frac{g(w)K}{\epsilon}\right]\!y\pm\frac{g(w)s_c a_s}{\epsilon}+\frac{g(w)s_c^2\beta}{\epsilon}\!\int\!y(t)dt\!\right)\end{array}
\end{equation}
where the $w$ in $g(w)$ depends on $y$ as
\begin{equation}\label{pr8}
w=\frac1{\alpha}y\mp s_ca_s\,t+s_c(K-s_c\beta)\!\int\!y(t)dt.
\end{equation}

\vspace{.1cm}
\noindent {\bf (3)}  For a Newtonian formulation $z''=F(t,z,z')/m$ of (\ref{eq1}) we have
\begin{equation}\label{pr9}
\frac{F}m=-s_c^2\alpha\beta\left( \frac{K}{s_c\beta}z'+\left[\frac1{\epsilon\beta}-1\right]z\pm a_s-\frac1{\epsilon}\!\int\!g(w)dw\right)
\end{equation}
where the $w$ in $g(w)$ depends on $z$ as
\begin{equation}\label{pr10}
w=-\frac1{s_c\alpha\beta}z'-\frac{K}{\beta}z\mp s_ca_st+s_c\!\int\!z(t)dt.
\end{equation} $\hfill \diamond$

\vspace{.15cm}
\noindent \emph{Comment 3.2.} All expressions of $F$ in (\ref{pr6}), (\ref{pr7}) and (\ref{pr9}) are  non-autonomous. 

\noindent \emph{Comment 3.3.} The fact that the system (\ref{eq68}) is jouncely Newtonian in all four variables is rather remarkable considering the fact that many well-known systems with three variales, i.e. those of Lorenz and R\"{o}ssler, are not even \emph{jerky Newtonian} in all three variables \cite{paper19}.

\section{The $rms$-equivalence of two-port $RL$ and $RC$ circuits}
Suppose now that the quantities in Fig.\ref{Fig2} are defined over one period. The general diagram in Fig.\ref{Fig2} can be represented for VCMR and CCMR in terms of the voltage, current, flux and charge variables $v$, $i$, $\psi$ and $q$, respectively, as shown in Figs.\ref{Fig11a} (VCMR) and \ref{Fig11b} (CCMR).

\begin{figure}[b!]
\begin{center}
\subfigure[The VCMR diagram]
{\label{Fig11a}\includegraphics*[height=1.7in,width=2.3in]{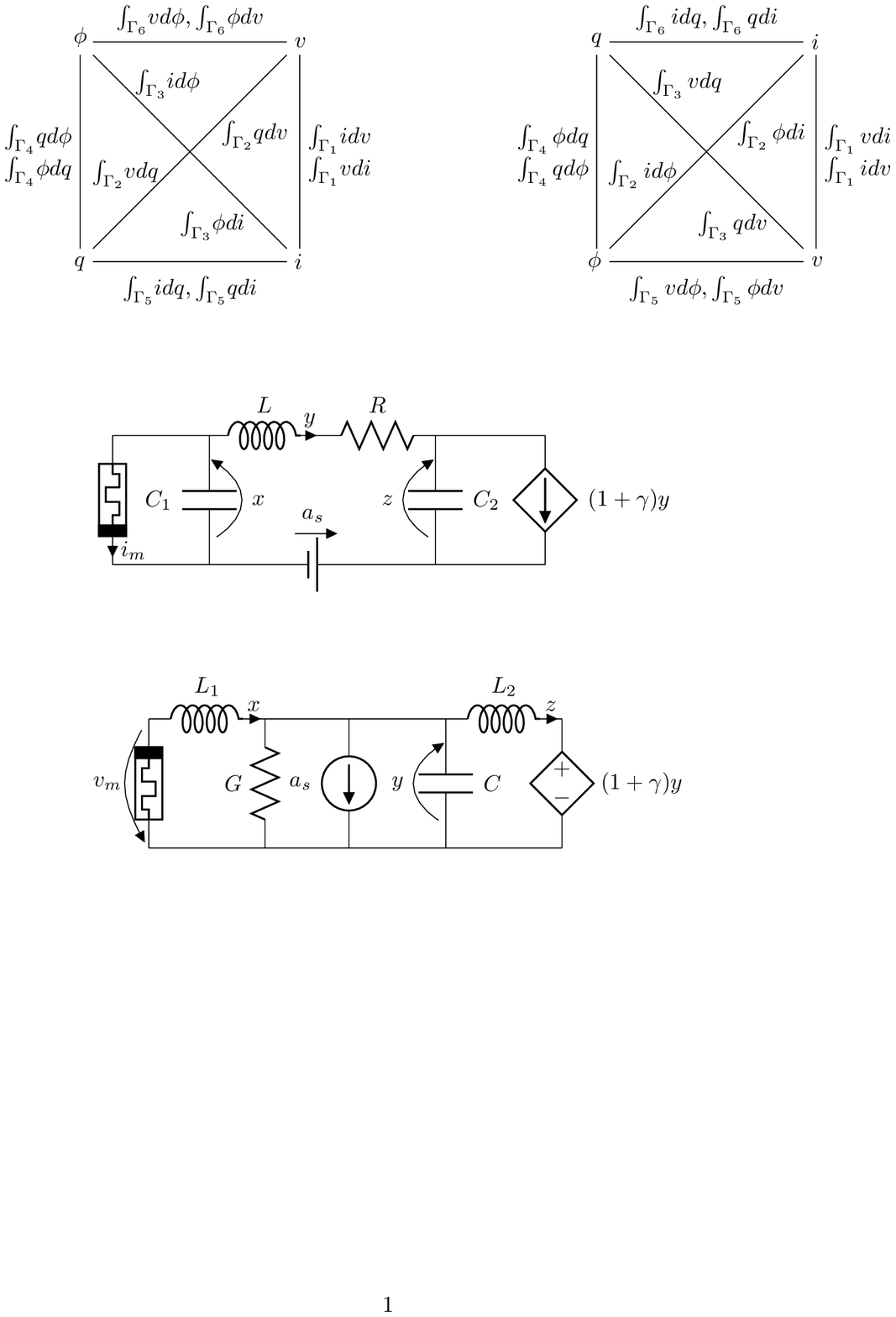}}
\subfigure[The CCMR diagram]
{\label{Fig11b}\includegraphics*[height=1.7in,width=2.3in]{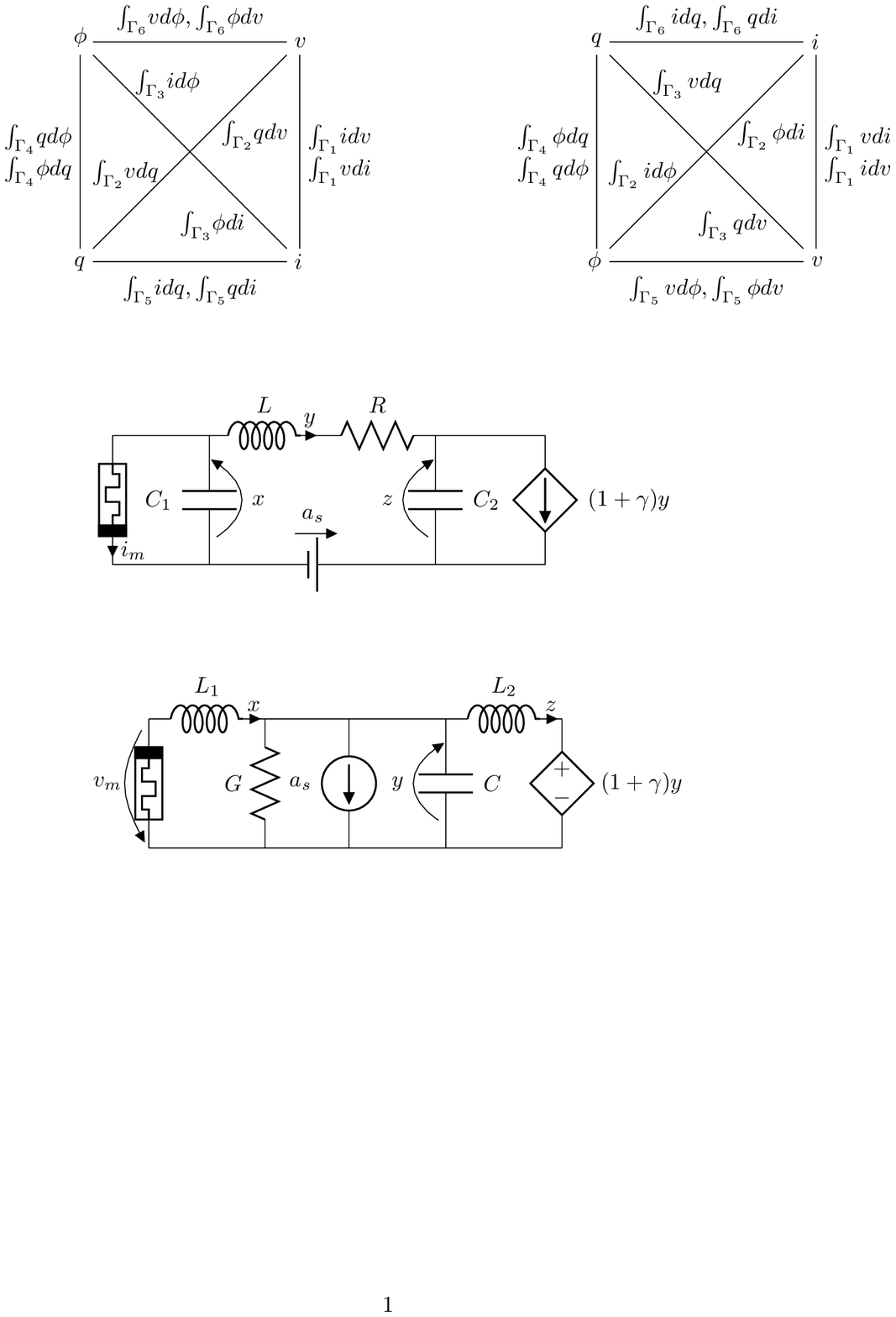}} 
\caption{Diagrams for VCMR (a) and CCMR (b).}
\end{center}
\end{figure}

Since the $\Gamma_i$, $i=1,\dots,6$, represent closed curves for $0\le t\le T$, where $T$ stands for the period, therefore for general curve $\Gamma$ we have $\int_{\Gamma}vd\phi=\int_0^Tv\frac{d\phi}{dt}dt=\int_0^Tv^2dt=Tv_{rms}^2$, where the root-mean-square value $v_{rms}=\sqrt{\frac{1}{T}\int_0^Tv^2dt}$. Also, $\int_{\Gamma}\phi dv=\phi v|_0^T-\int_0^Tvd\phi=-Tv_{rms}^2$ (since $\phi v|_0^T=0$). In a similar way one can prove that $\int_{\Gamma}idq=Ti_{rms}^2$ and $\int_{\Gamma}qdi=-Ti_{rms}^2$. This provides an interpretation of the one-period quantities on the top and bottom sides of the diagrams in Fig.\ref{Fig11a} and \ref{Fig11b}: the integrals are simply equal to the positive or negative of the period $T$ multiplied by the square  of the memristor's  voltage or current \emph{rms} values.

Similar analysis can be done for the one-period quantities represented by the integrals on the diagonals of the diagrams in Figs.\ref{Fig11a} and \ref{Fig11b}. For example, $\int_{\Gamma}vdq=\int_0^Tv\frac{dq}{dt}dt=\int_0^Tvidt=\mathcal{E}_C$, the memristor's one period (electric) energy. Also,  $\int_{\Gamma}id\phi=\int_0^Ti\frac{d\phi}{dt}dt=\int_0^Tivdt=\mathcal{E}_L$, the memristor's one period (magnetic) energy. 

Using the above $\mathcal{E}_C$, $\mathcal{E}_L$ and $v_{rms}$, $i_{rms}$ values  we can calculate   parallel $G$-$C$ (conductance-capacitance) and series $R$-$L$ (resistance-inductance) equivalent linear circuits  yielding the same \emph{rms} sinusoidal values as those obtained in the nonlinear memristor circuits. For the $G$-$C$ parallel circuit we have $GTv^2_{rms}=\mathcal{E}_C$, while for the $R$-$L$ series circuit we have $RTi^2_{rms}=\mathcal{E}_L$. From the first equation and the fact that $Tv^2_{rms}=\int_0^Tv^2dt$ we have \begin{equation}\label{eqG}
G=\frac{\mathcal{E}_C}{\int_0^T\!\!v^2dt}.
\end{equation}
Also, for the memristor's admitance we obtain $Y=\frac{i_{rms}}{v_{rms}}=\sqrt{\int_0^T\!i^2dt/\int_0^T\!v^2dt}$. Thus
\begin{equation}\label{eq56}
Y^2 = \left (\frac{\mathcal{E}_C}{\int_0^T\!v^2dt}\right )^2 + \left (\frac{2\pi}{T}C\right )^2=\frac{\int_0^T\!i^2dt}{\int_0^T\!v^2dt}
\end{equation}
from which we obtain
\begin{equation}\label{eq57}
C=\frac{T}{2\pi \int_0^T\!v^2dt}\sqrt{\int_0^T\!\!v^2dt \int_0^T\!\!i^2dt-\mathcal{E}_C^2 }.
\end{equation}
Thus, the values of $G$ and $C$ can be computed from (\ref{eqG}) and (\ref{eq57}), respectively. The period $T$ for the circuits in Fig.\ref{Rys2c} and \ref{Rys2d} can be estimated by the formulae derived in \cite{paper10}.

In an analogous way we can obtain an equaivalent $R$-$L$ series circuit. The emphasis is to have a one-to-one correspondence between our memristive circuits and their equivalent linear  parallel $G$-$C$ or series $R$-$L$ series to yield the same \emph{rms} values.

\section{Conclusions}
We have analyzed various oscillatory memristive circuits and proved their links to Newton's law, as the internal memristor's variables  satisfy the equation $w''=F(t,w,w')/m$ with a non-autonomous functions $F$ containing memory terms. For the memristive circuits with MMOs all dynamical variables satisfy Newton's law as shown in section 4. Once a Newton's law equation is obtained for various variables, we can transform such an equation into a jounce  scalar equation, with the term jounce has been borrowed from physics/mechanics, where it  means the second time derivative of the acceleration. The jounce equation for the memristive circuits with MMOs (Figs.\ref{Rys2c} and \ref{Rys2d}) has  been used to construct an equivalent jounce Newtonian  circuit in SPICE (Fig.\ref{Fig5}). Similar approach can be applied to Chua's circuits shown in Figs.\ref{Rys2a} and \ref{Rys2c}.

   The diagrams constructed in this paper (Figs.\ref{Fig2}, \ref{Fig11a} and \ref{Fig11b}) with various integral quantities can be interpreted as the power  (right sides of the diagrams in Figs.\ref{Fig11a} and \ref{Fig11b}), energy, \emph{rms} and action values. The later has the dimensions of [energy]$\times$[time], and its SI unit is  Joule}$\times$second. This is an interesting new (or rather forgotten in the circuit theory) quantity that further links memristors to physics and quantum mechanics throught the famous  
 Planck's constant. The Planck's constant has also the same unit  Joule$\times$second, as the action $\mathcal{A}$ does. The Planck's constant is used in the relationship between energy and frequency of an electromagnetic wave, known as the Planck-Einstein equation $E=h\nu$ (or $h=E/\nu=TE$), where $E$ is the energy of the charged atomic oscillator, $\nu$ is the frequency of an associated electromagnetic wave and $h$ is the Planck's constant.
Finally, we showed how to find linear parallel $G$-$C$ and series $R$-$L$ two-port circuits yielding the same \emph{rms} values as those in the memristive circuits. The assumption is that   the $G$-$C$ and $R$-$L$ circuits have sinusoidal inputs.

\end{document}